\documentclass[11pt]{llncs}
\usepackage{graphicx}
\usepackage{cite}
\usepackage{url}
\usepackage{times}
\usepackage{mathfont}
\def\log{\mathop{{\rm log}}}

\DeclareSymbolFont{AMSb}{U}{msb}{m}{n}
\DeclareSymbolFontAlphabet{\Bbb}{AMSb}
\def\Real{\ensuremath{\Bbb R}}

\mathcode`O="724F

\makeatletter
\def\hb@xt@{\hbox to }
\makeatother

\let\oldendproof\endproof
\def\endproof{\qed\oldendproof}

\begin{document}

\title{Dynamic Generators of Topologically Embedded Graphs} 

\author{David Eppstein}

\institute{UC Irvine,
Dept. of Information \& Computer Science\\
\email{eppstein@uci.edu}}

\maketitle   

\pagestyle{plain}

\begin{abstract}
We provide a data structure for maintaining an embedding of a graph on a surface
(represented combinatorially by a permutation of edges around each vertex)
and computing generators of the fundamental group of the surface, in amortized time
$O(\log n + \log g(\log\log g)^3)$ per update on a surface of genus $g$; we can also test orientability of the surface in the same time, and maintain the minimum and maximum spanning tree of the graph in time $O(\log n + \log^4 g)$ per update.  Our data structure allows edge insertion and deletion as well as the dual operations; these operations may implicitly change the genus of the embedding surface.
We apply similar ideas to improve the constant factor in a separator theorem for low-genus graphs, and to find in linear time a tree-decomposition of low-genus low-diameter graphs.
\end{abstract}

\section{Introduction}

In this paper we introduce a tool, the {\em tree-cotree decomposition}, for constructing generators of fundamental groups of surfaces, and apply it to several algorithmic problems for graphs embedded on surfaces.

The {\em fundamental group} of a surface measures the ability to transform one curve continuously into another.  Two closed curves $c_0$ and $c_1$ on a surface are {\em homotopic} if there exists a continuous map from the cylinder $[0,1]\times S^1$ to the surface such that each cylinder boundary $i\times S^1$ is mapped homeomorphically to curve $c_i$.  The elements of the fundamental group are equivalence classes of curves under homotopy. The identity element is the class of  curves  homotopic to the boundary of a disk, the inverse of a curve is formed by tracing the same curve backwards, and the concatenation of two curves forms their product in the group. For further exposition see any topology text, e.g.~\cite{Sti-93}.

Fundamental groups are often infinite, but can be specified finitely by a system of {\em generators} and {\em relations}.  The generators of a fundamental group are a system of closed curves with the property that any other curve can be generated (up to homotopy) by concatenations of closed curves in the system.
For instance, on a torus, two natural closed curves to use as generators are an equatoral circle $g_0$ and a longitudinal circle $g_1$; any other closed curve $c$ on the torus is homotopic to a curve $g_0^ig_1^j$ that winds $i$ times around the equator followed by $j$ turns around a longitude.  The relations of a fundamental group describe null-homotopic combinations of generators: in the torus, the relation $g_0g_1g_0^{-1}g_1^{-1}=1$ implies that the group is commutative.

In addition to their topological interest, fundamental groups and their generators have other applications.  Cutting a surface along its generator curves produces a planar surface
with holes at the cuts, and Erickson and Har-Peled~\cite{EriHar-SCG-02} have recently investigated the problem of finding short cutsets in nonplanar surfaces.
As we will see, this cutting technique allows one to apply planar graph
algorithms such as separator construction~\cite{AleDji-SJDM-96} and tree decomposition~\cite{Epp-Algo-00} to nonplanar graphs.  We will also show how to use
the generators to test orientability of surfaces and maintain minimum spanning trees
of graphs embedded in surfaces.

Subgroups of the fundamental group correspond to {\em covering spaces},
the most important of which is the {\em universal cover}.
In the example of the torus, the universal cover is a plane,
and the endpoints of curves $g_0^ig_1^j$ lift to a planar lattice.
If one chooses any pair of generators of the lattice, curves from the origin to these points
map to a pair of generators for the fundamental group of the torus.
Thus construction of generators is also closely related to lattice basis reduction,
Euclid's algorithm, and continued fractions.

\subsection{New Results}

We show the following results.

\begin{itemize}
\item We provide a simple dynamic graph data structure for maintaining a dynamic graph, embedded in a 2-manifold, subject to updates that insert or delete edges, changing the ordering of edges around each vertex and thus implicitly changing the underlying 2-manifold.  Our structure can also handle the dual operations of edge contraction and expansion, and maintains a set of generators for the fundamental group of the surface, as well as additional information such as the minimum spanning tree of the graph or the orientability of the surface, in amortized time $O(g\log n)$ per update where $g$ is the genus of the surface at the time of each update.

\item By combining our simple structure with ideas of separator based sparsification~\cite{EppGalIta-JCSS-96} and recent polylogarithmic algorithms for dynamic connectivity in general graphs~\cite{HolLicTho-JACM-01,Tho-STOC-00}
we further improve our time bounds, to $O(\log n + \log g(\log\log g)^3)$ per update for generators and orientability, and $O(\log n + (\log g)^4)$ per update for minimum spanning trees.

\item We improve by a factor of $\sqrt{2}$ the constant factor in the best previous separator theorem for bounded genus graphs~\cite{AleDji-SJDM-96}.

\item We provide an efficient algorithm for our previous nonconstructive result,
that genus $g$ graphs with diameter $D$ have a tree-decomposition with treewidth $O(gD)$~\cite{Epp-Algo-00}.
\end{itemize}

\subsection{Related Work}

The previous work most close to ours is on construction of canonical schemata and cutsets.

A {\em canonical schema} for a 2-manifold is a set of generators for its fundamental group,
having a single prespecified relation
($\prod g_{2i}g_{2i+1}g_{2i}^{-1}g_{2i+1}^{-1}$ for oriented surfaces,
$\prod g_i^2$ for unoriented surfaces).
The generators must have a common basepoint, but unlike the ones in our construction
they are not required to follow the edges of a given graph on the surface: they may pass across the interior of cells.  Vegter and Yap~\cite{VegYap-SCG-90}
showed that canonical schemata exist with total complexity $O(ng)$ where $g$ is the surface genus; this bound is tight in the worst case, and they also showed that a canonical schema can be found in time $O(ng)$.  The generators we find can have the same $\Theta(gn)$ total complexity, but we can find an implicit representation of the generators in linear time.
However, the generators we find are not necessarily in canonical form.

Recently, Erickson and Har-Peled~\cite{EriHar-SCG-02} studied the problem of
finding a {\em cutset}, that is, a set of edges the complement of which forms a topological disk on the given surfaces.  Erickson and Har-Peled found algorithms for computing minimum-length cutsets in polynomial time for bounded-genus embeddings, but showed the problem to be NP-complete in the case of unbounded genus.  In the applications to tree decomposition and separator theorems, we use our techniques to find cutsets with a guaranteed bound on the total length, but our cutsets may be far from the minimum possible length.

Planar separator theorems have been long studied~\cite{LipTar-SJAM-79}
and there has been much work on improved constants for such theorems~\cite{SpiTen-SCG-96}.
More generally, for graphs embedded on genus $g$ surfaces,
a separator with $O(\sqrt{gn})$ vertices can be found in $O(n)$ time~\cite{AleDji-SJDM-96}.
Specifically, Aleksandrov and Djidjev~\cite{AleDji-SJDM-96} showed that, for any $\epsilon>0$,
one can find a set of $\sqrt{(16g+O(1/\epsilon))n}$ vertices the removal of which partitions
the graph into components of size at most $\epsilon n$.  Our techniques improve the constant factors in this result.  We also apply similar ideas to the computation of tree-decompositions
for low-diameter low-genus graphs; our previous work~\cite{Epp-Algo-00} proved the existence of such decompositions but did not provide an efficient algorithm for finding them.

The main inspiration for this paper was previous work on dynamic connectivity and minimum spanning trees in embedded planar graphs~\cite{EppItaTam-Algs-92} which worked
by decomposing the graph into two subgraphs, a minimum spanning tree and a maximum spanning cotree.  We use a similar decomposition here, however for surfaces of nonzero genus
we end up with leftover edges that belong neither to the tree nor to the cotree.
These extra edges are what we use to form the generators of the surface.
We also combine the dynamic graph algorithms from that previous work, with
a technique for maintaining small representative subgraphs
\cite{EppGalIta-JCSS-96} and more recent methods of Holm et al.~\cite{HolLicTho-JACM-01}
to achieve time bounds that match the best of these dynamic graph algorithms.

\section{Representing an Embedding}
\label{rep}

\begin{figure}[t]
\centering
\includegraphics[width=5in]{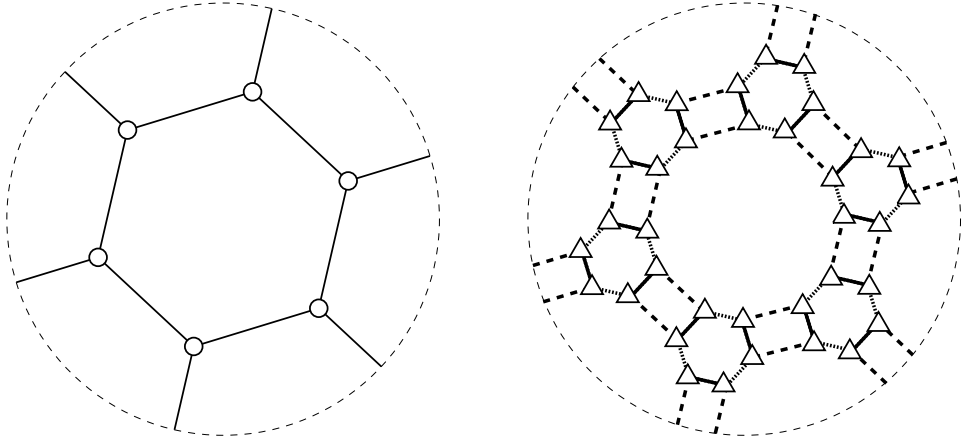}
\caption{Left: an embedding of $K_{3,3}$ in the projective plane.  The edges drawn crossing the dashed circle connect to the opposite side of the circle; there is one hexagonal cell in the center surrounded by three quadrilaterals that form a M\"obius strip.  Right: a gem representation of the embedding.  The flags of the embedding are shown as triangles; the dashed edges represent connections of type $S_V$, the dotted edges represent connections of type $S_E$, and the solid edges represent connections of type $S_C$.}
\label{fig:gem}
\end{figure}

We consider {\em maps}: graphs embedded on two-dimensional manifolds, in such a way that the cells of the embedding are disks.  We do not require that our graphs be simple.  Such an embedding on an orientable manifold can be represented by specifying a clockwise ordering of the edges around each vertex~\cite[Sedction 3.2.4]{GroTuc-01}.  However for our purposes it is convenient to use a somewhat more general representation, the graph-encoded map~\cite{BonLit-95,Lin-JCTB-82}, which also allows unorientable embeddings.  We now briefly describe this representation.

Suppose we are given a map, such as the embedding of $K_{3,3}$ shown on the left of Figure~\ref{fig:gem}.  A {\em flag} of the map is defined to be the triple formed by a mutually adjacent vertex, edge, and cell of the map.  For a flag $F$, let $V(F)$ denote the vertex, $E(F)$ denote the edge, and $C(F)$ denote the cell.  We define three operations that change one flag $F$ into another:
\begin{itemize}
\item $S_V(F)$ is a flag formed by the edge and cell of $F$ but with a different vertex, found at the other endpoint of $E(F)$.
\item $S_E(F)$ is a flag formed by the vertex and cell of $F$ but with a different edge, found by tracing around the boundary of $C(F)$ from $E(F)$ in the direction of $V(F)$.
\item $S_C(F)$ is a flag formed by the vertex and edge of $F$ but with a different cell,
the cell that is the next one clockwise or counterclockwise around $V(F)$ from $C(F)$ in the direction of $E(F)$. 
\end{itemize}

Each of these operations produces a unique well-defined flag, and the neighboring relation between flags is symmetric: $S_V(S_V(F))=S_E(S_E(F))=S_C(S_C(F))=F$.
In a computer, one can represent these flags as objects in an object-oriented language, and store with each flag a pointer or reference to its three types of neighbor.
Graphically, we view the flags as vertices in a graph, and the operations as unoriented edges,
connecting each flag to its neighbors $S_V(F)$, $S_E(F)$, and $S_C(F)$, as shown on the right of Figure~\ref{fig:gem}.  Although we see this gem as drawn in a projective plane,
the gem representation of an embedding is purely graph-theoretic and independent of any drawing of the gem.  The edges of this graph should be viewed as having three colors, representing the type of operation generating the edge; in the figure these are shown as dashed, dotted, and solid, respectively.

For each vertex of the map, with $d$ adjacent edges, the resulting gem has a $2d$-cycle with edges alternating between types $S_E$ and $S_F$.  For each edge of the map, the gem has a quadrilateral with edges alternating between types $S_V$ and $S_C$.  And for each $k$-gon cell of the map, the gem has a $2k$-cycle with edges alternating between types $S_V$ and $S_E$.
In this way, every map generates a 3-regular 3-edge-colored graph in which the $S_V$-$S_C$ cycles all have four edges.  Conversely, given any such graph, we can recover the original map by 
forming a $k$-gon cell for each $2k$-cycle of edges of types $S_V$ and $S_E$ and by gluing these cells together into a manifold whenever the corresponding edges of types $S_V$ are part of a four-cycle of types $S_V$ and $S_C$.

Any map $\mathcal M$ has a {\em dual map} $\mathcal M^*$ on the same surface, formed by creating a dual vertex $f^*$ within each face $f$ of the primal map, and creating a dual edge $e^*$ for every primal edge~$e$, so that if $e$ is adjacent to two faces $f_1$ and $f_2$, then $e^*$ connects $f^*_1$ and $f^*_2$ by a path
that crosses $e$ once and crosses no other primal or dual edge.  In the gem representation, the dual can be formed very simply, by reversing the roles of the gem edges of types $S_V$ and $S_C$.

In our dynamic graph data structures, we will augment the gem representation by balanced binary trees for the cycles corresponding to each vertex and cell of the map; in this way we can quickly look up which vertex's or cell's cycle contains a given edge of the gem, and whether two edges on the same cycle are in the same or opposite orientations around the cycle.

\section{Trees and Generators}

A {\em spanning tree} of a graph or map is just a tree formed by some subset of the edges of the graph that incorporates all the vertices of the graph.  If $C^*$ is a spanning tree of the dual of a map,
we call $C=\{e\mid e^*\in C^*\}$ a {\em spanning cotree} of the map.  If the edges of a map are given weights, the weight of a tree or cotree is defined to be the sum of the weights of its edges.

\begin{lemma}
\label{minmax}
Let a map $\mathcal M$ be given, with distinct weights on each of its edges.
Then the minimum weight spanning tree of $\mathcal M$ and
the maximum weight spanning cotree of $\mathcal M$ are disjoint.
\end{lemma}

\begin{proof}
Let $e$ be an edge in the minimum spanning tree of $\mathcal M$.
Then removing $e$ from the minimum spanning tree results in a forest $F$ with two connected components; let $F_1$ be one of those components.  Then the sequence of faces of $\mathcal M$ surrounding $F_1$ forms a (possibly non-simple) cycle in $\mathcal M^*$ consisting of the duals of all edges connecting $F_1$ to $F_2$.  Since $e$ belongs to the minimum spanning tree, $e^*$ must be the shortest edge in this dual cycle, and so cannot belong to the maximum spanning tree of $\mathcal M^*$.
\end{proof}

In the special case of planar graphs, the minimum spanning tree and maximum spanning cotree form a partition of all the edges of the graph~\cite{EppItaTam-Algs-92}.  We define a {\em tree-cotree partition} of $\mathcal M$ to be a triple $(T,C,X)$ where $T$ is a spanning tree of $\mathcal M$, $C$ is a spanning cotree of $\mathcal M$, and the three sets $T$, $C$, and $X$ are disjoint and together include all edges of $\mathcal M$.  In particular, if $T$ is the minimum spanning tree and $C$ is the maximum spanning cotree, then $(T,C,E(\mathcal M)\setminus(T\cup C))$ is a tree-cotree decomposition.  By assigning weights appropriately, we can use Lemma~\ref{minmax} to find a tree-cotree decomposition involving any given tree $T$ or cotree $C$.  More generally, if $T$ and $C^*$ are forests such that $T$ and $C$ are disjoint, we can assign weights to make $T$ become part of the minimum spanning tree and $C$ become part of the maximum spanning cotree, extending $T$ and $C$ to a tree-cotree decomposition.

We now show how these decompositions are connected to fundamental groups of surfaces.
The {\em fundamental group} of a space $S$ is most commonly defined relative to some base point $x_0\in S$, but for suitable spaces (including the 2-manifolds considered here) it is independent of the choice of this base point.  Define a {\em loop} to be a continuous function $f:[0,1]\mapsto S$ satisfying $f(0)=f(1)=x_0$, and define two loops $f_0$ and $f_1$ to be {\em homotopy equivalent} if there exists a continuous function $f:[0,1]^2\mapsto S$ such that $f(x,i)=f_i(x)$ and $f(0,i)=f(1,i)=x_0$.
Define the {\em product} of two loops $f_0$ and $f_1$ to be the loop
$$f(x)=\left\{{f_0(2x)\mbox{\qquad for $0\le x\le 1/2$}\atop
f_1(2x-1)\mbox{\qquad for $1/2\le x\le 1$}}\right..$$
Then it can be shown that homotopy equivalence is an equivalence relation,
and that the homotopy equivalence classes of loops form a group, the {\em fundamental group},
under homotopy equivalence.  In this group, inverses can be found by reversing the direction of a loop: if $f$ is a loop, $g(x)=f(1-x)$ is a loop inverse to $f$.
A set of loops is said to {\em generate} the fundamental group if every other equivalence class of loops can be reached by products of these generators and their inverses.

If we are given a rooted tree $T$ and oriented edge $e\notin T$, with the base point $x_0$ of the fundamental loop at the root of $T$, we can define a loop
${\rm loop}(T,e)$ by following a path in $T$ from $x_0$ to the head of $e$, then traversing edge $e$ to its tail, and finally following a path in $T$ from that tail back to $x_0$.
When considering these loops as generators of the fundamental group, the orientation of $e$ is unimportant, since reversing the edge merely produces the inverse group element.

\begin{lemma}
\label{gen}
Let $(T,C,X)$ be a tree-cotree decomposition of map $\mathcal M$.
Then the loops $\{{\rm loop}(T,e)\mid e\in X\}$ generate the fundamental group of the surface on which $\mathcal M$ is embedded.
\end{lemma}

\begin{proof}
Contract the edges in $T$ into the root $x_0$ of $T$, while leaving the surface unchanged outside of a small neighborhood of $T$; this contraction does not change the fundamental group of the surface.
In the contracted image, the loops from the statement of the lemma are each contracted into a single edge connecting $x_0$ to itself.  Decompose the surface into a set $S_1$ consisting of a small neighborhood of the set of these contracted loops, and a set $S_2$ consisting of the faces of the map and the cotree edges in $C$.
Then, by the Seifert -- Van Kampen theorem~\cite[Section 3.4]{Sti-93} the fundamental group of the overall surface can be formed by combining the generators and relations of the fundamental groups of these two pieces, with additional relations describing the way these two pieces fit together.  $S_1$ can be contracted to a graph with one vertex and $|X|$ self-loops, which has a fundamental group with each of these loops as generators and no relations.
$S_2$ is topologically a disk (since it consists of a collection of disk faces glued together in the pattern of a tree) so it has a trivial fundamental group.  The fundamental group of the overall surface, then, is generated by the loops described in the statement of the lemma, with a single relation formed by the concatenation of these loops around the boundary of $S_2$.
\end{proof}

Of course, for any tree $T$, the loops
$\{{\rm loop}(T,e)\mid e\notin T\}$ also generate the fundamental group, but in general this
will form a much larger system of generators.  The advantage of the system of generators provided by Lemma~\ref{gen} is its small cardinality, proportional to the genus of the surface.

The same tree-contraction argument used in Lemma~\ref{gen} also shows that the edges in the  loops $\{{\rm loop}(T,e)\mid e\in X\}$ form a cutset.

\section{Dynamic Generators}
\subsection{Update Operations}

\begin{figure}[t]
\centering\includegraphics[width=6in]{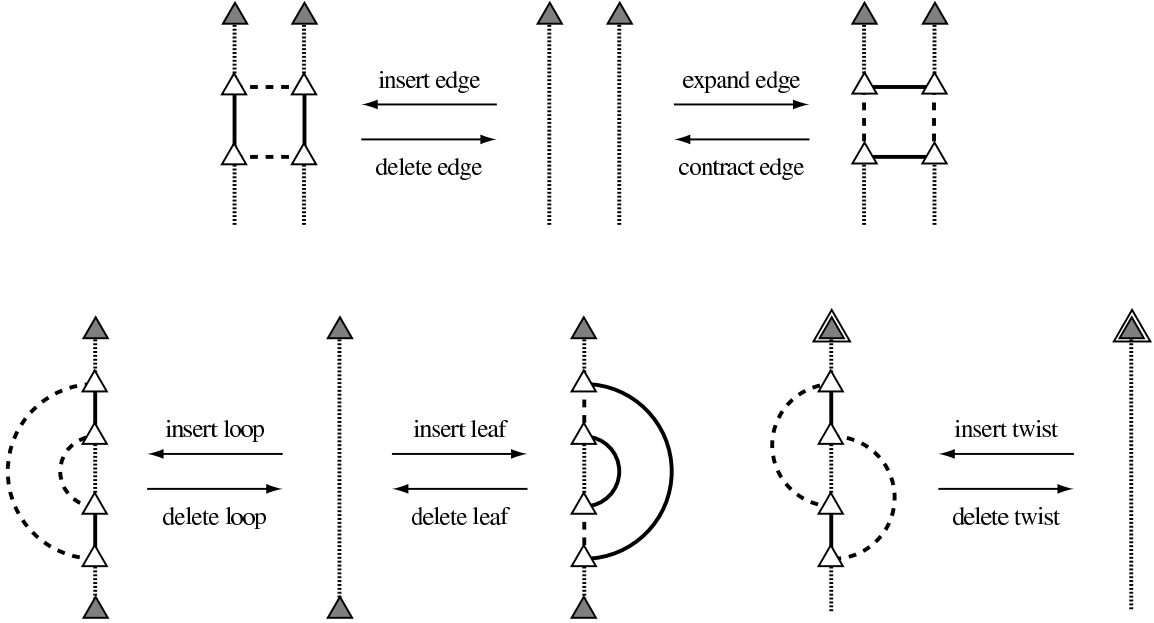}
\caption{Effects of edge insertion and deletion and their duals on the gem representation.}
\label{fig:operations}
\end{figure}

Following \cite{EppItaTam-Algs-92}, we allow four types of update operation: insertion
of a new edge between two vertices, deletion of an edge, dual edge insertion (expansion), and dual edge deletion (contraction).  The effect of these updates on the gem representation is shown in Figure~\ref{fig:operations}.

The location of an edge insertion can be specified by a pair of gem vertices $p$ and $q$, shown as shaded triangles in the figure.
We then subdivide the two gem edges $S_E(p)$ and $S_E(q)$ into paths
of $S_E$ and $S_C$ edges, preserving the connectivity of the $S_E$-$S_C$ cycles
representing vertices in the gem, and
create a new graph edge (represented in the gem by a 4-cycle of $S_V$ and $S_C$ edges)
using the $S_C$ edges of the subdivided paths, in such a way that the new gem edges $S_E(p)$ and $S_E(q)$ are adjacent to a common edge of type $S_V$.

In the most typical case of an edge insertion (Fig.~\ref{fig:operations}, upper right), the two edges $S_E(p)$ and $S_E(q)$ are distinct; each of these edges is subdivided into a three-edge path, and the two new vertices on each path are connected in pairs by new $S_V$ edges.
We distinguish three subcases of this operation, depending on which cells of the map contain the endpoints of the inserted edge:

\begin{itemize}
\item In a {\em cell-merging insertion}, the two endpoints of the inserted edge are attached into
distinct cells of the map; that is, the two gem edges $S_E(p)$ and $S_E(q)$ belong
to distinct $S_E$-$S_V$ cycles.  The effect of the insertion is to split each of these two cycles
into paths, and to splice these two paths together into a single cycle via the newly created
$S_V$ edges.  We are increasing the number of edges and decreasing the number of cells, so the surface's genus increases.  Topologically, this can be viewed as gluing a handle between the two cells, and routing the new edge along this handle.

\item In a {\em cell-splitting insertion}, the two endpoints of the inserted edge belong to a single cell of the map, and if the $S_E$-$S_V$ cycle representing this cell is consistently oriented, then $S_E(p)$ and $S_E(q)$ both point towards or both point away from their endpoints $p$ and $q$ respectively.  The effect of the insertion is to split the $S_E$-$S_V$ cycle into two paths,
and then reconnect the paths into two cycles, forming two cells connected across the new edge.
We are increasing both the number  of edges and the number of cells, so the genus and
the topology of the surface are preserved.  This was the only type of edge insertion allowed in~\cite{EppItaTam-Algs-92}.

\item In a {\em cell-twisting insertion}, the two endpoints of the inserted edge belong to a single cell of the map, but $S_E(p)$ and $S_E(q)$ are not consistently oriented.  The effect of the insertion is to split the $S_E$-$S_V$ cycle into two paths, and then reconnect the paths into a single cycle,
with the orientation of the two paths reversed from their previous situation.
We are increasing the number of edges while preserving the number of cells, so the genus
increases.  Topologically, this can be viewed as gluing a crosscap into the cell,
and routing the new edge through it.
\end{itemize}

Each of these types of insertion has a complementary deletion, specified by the edge to be deleted, which we call a
{\em cell-splitting deletion}, a {\em cell-merging deletion}, or an {\em untwisting deletion}
respectively.
In addition, we allow the dual operations to insertions and deletions, as shown on the upper right of the figure;
dual deletion is more familiar as the operation of edge contraction.
The dual of a cell-merging insertion merges two vertices of the graph while increasing the genus, the dual of a cell-splitting insertion splits a vertex
into two adjacent vertices, and the dual of a cell-twisting insertion changes the order of the edges and cells around a vertex.  We do not describe these in detail because our data structure will be self-dual, so that each dual operation is performed exactly as the primal operation substituting $S_V$ for $S_C$ and vice versa.

Finally, it is possible to specify an insertion in which $S_E(p)$ and $S_E(q)$ are not distinct.
If $p\neq q$ while $S_E(p)=S_E(q)$ (Fig.~\ref{fig:operations}, lower left), we subdivide $S_E$ into a path of five edges, and connect the four new gem vertices into an $S_V$-$S_C$ cycle
in such a way that adjacent vertices of the cycle are connected to $p$ and $q$.
This can be viewed as a special case of a cell-splitting insertion in which the two endpoints
of the inserted edge are placed next to each other at the same vertex, creating a one-sided cell contained within the previously existing cell.  The dual of this operation adds a new degree-one vertex to the graph.

If $p=q$, we perform an edge insertion by again subdividing $S_E$ into a path of five edges, and connecting the four new gem vertices into an $S_V$-$S_C$ cycle in such a way that opposite vertices of the cycle are connected to $p$ and the other endpoint of $S_E(p)$.  Edge insertion with $p=q$ and dual edge insertion with $p=q$ produce the same result, shown in the lower right of Fig.~\ref{fig:operations}.
This can be viewed as a special case of a cell-twisting insertion in which the two endpoints
of the inserted edge are placed next to each other at the same vertex.

\subsection{Simple Dynamic Generators}

We now describe a simple scheme for maintaining generators and minimum spanning trees in dynamic maps, based on the dynamic planar graph algorithms of Eppstein et al.~\cite{EppItaTam-Algs-92}.  We will later show how to improve the time bounds for this scheme at the expense of some additional complexity.

\begin{theorem}
\label{simpledyn}
Let $\mathcal M$ be a map, with distinct edge weights, changing according to the update operations described in the previous section as well as by changes to the weights of the edges.
Then in time $O(g\log n)$ per update we can maintain a tree-cotree decomposition $(T,C,X)$, where $T$ is the minimum spanning tree, $C$ is the maximum spanning cotree,
and $g$ is the genus of the map at the time of the update operation.
\end{theorem}

\begin{proof}
We represent $T$ by Sleator and Tarjan's dynamic tree data structure~\cite{SleTar-JCSS-83}, modified as in~\cite{EppItaTam-Algs-92} to support edge contraction and expansion operations as well as the more usual edge insertions and deletions.  We use a similar data structure to represent the dual tree $C^*$.  With this data structure, we can find in logarithmic time the longest edge on any path of $T$, or the shortest edge on any path of $C^*$.   We can also test in logarithmic time whether a pair of edges $\{e,e'\}$ forms a {\em swap}; that is, whether $T\triangle \{e,e'\}$ (where $\triangle$ denotes the set-theoretic symmetric difference) is another spanning tree of $T$.
We also maintain the tree data structures discussed in Section~\ref{rep} for the adjacency lists of each vertex and face of $\mathcal M$, which allow us to classify the type of an insertion or deletion in logarithmic time.

To increase the weight of an edge $e$ in $T\cup X$, we first test whether the increased weight causes the edge to move to the maximum cotree $C$, by finding the shortest edge $e'$ on the dual path connecting the endpoints of $e^*$, and testing whether the dual swap $\{e,e'\}$ increases the weight of $C$.  If so, we add $e$ to $C$ and move $e'$ to $X$.  We then search $X$ for the minimum weight edge $e''$ forming a swap $\{e,e''\}$ in $T$, and testing whether this swap causes an improvement in the weight of $T$.  If it does, we add $e''$ to $T$ and move $e$ to $X$.
Lemma~\ref{minmax} shows that, after these steps, we must again have a tree-cotree partition.
To decrease the weight of an edge $e$ in $C\cup X$, we perform a similar sequence of steps on the dual tree-cotree-partition $(C^*,T^*,X^*)$.

We now describe the effect on our data structure of the various insertion and deletion operations.
The edge expansions and contractions are performed by very similar sequences of steps in the dual map.
\begin{itemize}
\item
To perform a cell-merging insertion of an edge $e$, we find the lightest edge $e'$ on the dual path in $C^*$ connecting the merged cells, and place both $e$ and $e'$ in $X$.
We then perform swaps similar to those for a weight change to determine whether $e$ should move into $T$ or $C$.  To perform a cell-splitting deletion of $e$, we search $X$ for the heaviest edge $e'$ connecting the two components of $C^*$ formed by the split, and move $e'$ into $C$.  

\item
To perform a cell-splitting insertion of an edge $e$, we add $e$ to $X$, search $X$ for the heaviest edge $e'$ connecting the two components of $C^*$ formed by the split, and perform swaps similar to those for a weight change to determine whether to move $e$ into $T$ or $C$.
To perform a cell-merging deletion of $e$, we increase the weight of $e$ until it belongs to $C$,
then remove it from the graph and merge its two dual endpoints.

\item
To perform a cell-twisting insertion of an edge $e$, we simply add $e$ to $X$, search for the best swap $\{e,e'\}$ with $e'\in T$, and perform this swap if it improves the weight of $T$.
The inserted edge $e$ can not belong to $C$ since $e^*$ connects a dual vertex to itself.
To perform an untwisting deletion, we test whether $e$ belongs to $T$, and if so swap it with the best replacement $e'\in X$ before removing it.
\end{itemize}

Each of these operations maintains the desired partition via a sequence of $O(g)$ dynamic tree queries and updates, so the total time per change to $\mathcal M$ is $O(g\log n)$.
\end{proof}

If we wish to maintain a tree-cotree decomposition of a graph without weights, we can choose weights for the edges arbitrarily.

\subsection{Improved Dynamic Generators}

The bottleneck of Theorem~\ref{simpledyn} is the search for swaps in graphs $T\cup X$ or $C^*\cup X^*$; performing this search by testing each edge in $X^*$ takes time $O(g\log n)$.
We can improve this by using the method of separator based sparsification~\cite{EppGalIta-JCSS-96} to replace $T$ and $C^*$ by contracted trees $T'$ and $C'$ that match the $O(g)$ size of $X$, and then applying a general dynamic graph connectivity algorithm~\cite{HolLicTho-JACM-01,Tho-STOC-00} to find the swaps in the smaller graphs $T'\cup X$ or $C'\cup X^*$.

Specifically, we form $T'$ from $T$ by repeated edge contractions, forming a sequence of trees $T_0=T, T_1, \ldots T_k=T'$.  If $T_i$ contains a vertex with degree one that is not an endpoint of an edge in $X$, we delete that vertex and its adjacent edge to form $T_{i+1}$.  Otherwise, if $T_i$ contains a vertex with degree two that is not an endpoint of an edge in $X$, let $e$ and $e'$ be
the two edges incident to that vertex; we form $T_{i+1}$ by contracting whichever of $e$ and $e'$ has the smaller weight.  The construction of $C'$ from $C^*$ is essentially the same except that we contract the larger weight of two edges incident to any degree two vertex.

\begin{lemma}
The trees $T'$ and $C'$ described above have $O(|X|)$ edges, and do not depend on the order in which the contractions are performed.
\end{lemma}

\begin{proof}
Consider the set of paths in $T$ connecting endpoints of $X$.  The union of these paths forms a subtree $T''$ of $T$.  Then an alternative description of $T'$ is that it consists of one edge for each path of degree-two vertices in $T''$, with the weight of this edge equal to the maximum weight of an edge in the path.  Specifically, each deletion of a degree one vertex removes an edge that is not in $T''$, and each contraction of a degree two vertex reduces the length of one of the paths in $T''$ without changing the maximum weight of an edge on the path.  This proves the assertion that $T'$ is independent of the contraction order; it has $O(|X|)$ edges because it has at most $|X|$ leaves and no internal degree-two vertices.
\end{proof}

\begin{lemma}
Trees $T'$ and $C'$ can be updated after any change to $T$ or $C^*$ in time $O(\log n)$ per change.  Each change to $T$, $C$, or $X$ causes $O(1)$ edges to change in $T'$ and $C'$.
\end{lemma}

\begin{proof}
We use the following data structures for $T$: First, we use a dynamic tree data structure~\cite{SleTar-JCSS-83} that can find the maximum weight edge on any path of $T$, and that can determine for any three query vertices of $T$ which vertex forms the branching point between the paths connecting the three vertex.  Second, we maintain an Euler tour data structure~\cite{HenKin-JACM-99} that can find, for a query vertex in $T$, the one or two nearest vertices that belong to $T'$.  Whenever a change to $X$ adds a vertex $v$ to the set of endpoints in $X$, we must also add $v$ to $T'$; we do this by querying the Euler tour data structure to find the edge or vertex of $T'$ at which the branch to $v$ should be added, querying the dynamic tree data structure to find the correct branching point on that edge, giving the overall graph structure of the new tree $T'$, and finally querying the dynamic tree data structure again to find the weights of the changed edges in $T'$.  To remove a vertex from the endpoints in $X$, we simply contract one or two edges as necessary in $T'$.  To perform a swap in $T$, we use the data structures as described above to make all swap endpoints part of $T'$, perform the swap, and then contract edges to remove unnecessary vertices from the changed $T'$.  The updates to $C'$ are similar.
\end{proof}

\begin{lemma}
Let $e$ be an edge of $T$ that was not contracted in forming $T'$, and let $e'$ be an edge of $X$.
Then $\{e,e'\}$ is a swap in $T$ if and only if it is a swap in $T'$.
\end{lemma}

\begin{lemma}
Let $e'$ be an edge in $X$, and let $e$ be the heaviest edge forming a swap $\{e,e'\}$ in $T$.
Then $e$ is not contracted in forming $T'$.
\end{lemma}

\begin{theorem}
Let $\mathcal M$ be a map, with distinct edge weights, changing according to the update operations described in the previous section as well as by changes to the weights of the edges.
Then in amortized time $O(\log n+(\log g)^4)$ per update we can maintain a tree-cotree decomposition $(T,C,X)$, where $T$ is the minimum spanning tree, $C$ is the maximum spanning cotree,
and $g$ is the genus of the map at the time of the update operation.
\end{theorem}

\begin{proof}
The overall algorithm is the same as the one in Theorem~\ref{simpledyn}, but
we use the data structures described above to maintain trees $T'$ and $C'$,
and use the dynamic minimum spanning tree data structure of Holm et al~\cite{HolLicTho-JACM-01} to find the edge $e'\in X$ forming the best swap for any edge
$e\in T$ or $e^*\in C^*$.
The time bound follows from plugging the $O(|X|)$ bound on the number of edges in $T'\cup X$ and $C'\cup X^*$ into the time bound for the minimum spanning tree data structure, and using this bound to replace the $O(g\log n)$ time for finding replacement edges in Theorem~\ref{simpledyn}.
\end{proof}

The final improvement to our bounds comes from the observation that, if we only wish to maintain a tree-cotree decomposition for a graph without weights, then we can replace the dynamic minimum spanning tree data structures for $T'\cup X$ and $C'\cup X$ with any dynamic connectivity data structure that maintains a spanning tree of these two dynamic graphs and changes the tree by a single swap per update to the graphs.
The connectivity structure of Holm et al~\cite{HolLicTho-JACM-01}, as improved by Thorup~\cite{Tho-STOC-00}, is a suitable replacement.

\begin{theorem}
Let $\mathcal M$ be a map, with distinct edge weights, changing according to the update operations described in the previous section as well as by changes to the weights of the edges.
Then in amortized time $O(\log n+\log g(\log\log g)^3)$ per update we can maintain a tree-cotree decomposition $(T,C,X)$, where $g$ is the genus of the map at the time of the update operation.
\end{theorem}

\subsection{Orientability}

The results above are nearly sufficient to determine the surface underlying the dynamic map $\mathcal M$, since it is known that 2-dimensional closed surfaces are classified by two quantities: their genus (which we can calculate from $|X|$) and the {\em orientability} of the surface.
We can compute the orientability from the following simple result, which follows from the fact that the loops generated by edges in $X$ form a cutset:

\begin{lemma}
Let $\mathcal M$ have a tree-cotree decomposition $(T,C,X)$.  Then
The surface on which $\mathcal M$ is embedded is unorientable
if and only if every sufficiently small neighborhood of at least one of the loops $\{{\rm loop}(T,e)\mid e\in X\}$ is homeomorphic to a M\"obius strip.
\end{lemma}

\begin{theorem}
Let $\mathcal M$ be a map, with distinct edge weights, changing according to the update operations described in the previous section as well as by changes to the weights of the edges.
Then in amortized time $O(\log n+\log g(\log\log g)^3)$ per update we can determine after each update whether the surface underlying the map is orientable.
\end{theorem}

\begin{proof}
For any path in $\mathcal M$, we can form a sufficiently small neighborhood of the path by gluing together a collection of disks, where each disk has as its boundary the gem cycle representing a vertex or edge in the path.
We choose an arbitrary clockwise orientation for the gem cycle representing each vertex in $\mathcal M$, and augment the Sleator-Tarjan dynamic tree data structure used to maintain the tree $T$ in the tree-cotree decomposition $(T,C,X)$, by storing an extra bit per tree edge
that is zero when the edge's preserves the chosen orientations of its two endpoints and one when it reverses these orientations; we can use the dynamic tree data structure to determine the parity of the sum of these bits along any path in $T$.  After each change to the tree-cotree decomposition,
we use this data structure to determine whether any $e\in X$ forms a loop with odd total parity (including the parity of $e$ itself).  If any odd loop is found, the map is unorientable; if all loops are even, the map is orientable.
\end{proof}

\section{Separator Theorems}

Since Lipton and Tarjan's discovery of the planar separator theorem~\cite{LipTar-SJAM-79}, separator theorems have become widely used in algorithms, and there has been much work on improved constants for these theorems.
Formally, we define an $\epsilon$-separator of an $n$-vertex graph $G$ to be a subset $S$ of the vertices of $G$ such that each connected component of $G\setminus S$ has at most $\epsilon n$ vertices.
For graphs embedded on a genus-$g$ orientable surface, Aleksandrov and Djidjev~\cite{AleDji-SJDM-96}
showed that there always exists an $\epsilon$-separator
with $|S|\le\sqrt{(16g+O(1/\epsilon))n}$.
We combine our tree-cotree decomposition idea with the Lipton-Tarjan method of partitioning the breadth first search tree into levels (also used by Aleksandrov and Djidjev) to improve this result and extend it to unorientable surfaces.

By Euler's formula, if $(T,C,X)$ is a tree-cotree decomposition of a map on an orientable surface, then $|X|=2g$ where $g$ is the genus of the surface on which $\mathcal M$ is embedded.
If the surface is unorientable, then $|X|=g$.  Because of this confusion between orientable and unorientable surfaces, we prefer to use the {\em Euler characteristic} $\chi(\mathcal M)=|X|-2$
in stating our results.  In terms of $\chi$, Aleksandrov and Djidjev's result is that there exists a separator with $|S|\le\sqrt{(8\chi+O(1/\epsilon))n}$.  We improve this by a factor of $\sqrt{2}$, to
$|S|\le\sqrt{(4\chi+O(1/\epsilon))n}$.

Let $T$ be a breadth first search tree of the given graph, and form a tree-cotree decomposition $(T,C,X)$.
For any pair of integers $0\le i<k$, define vertex sets $L_{i,k}$ and $S_{i,k}$ as follows:
$L_{i,k}$ consists of all vertices $v$ whose depth $d(v)$ in $T$ satisfies
$d(v)\equiv i$ (mod $k$).
$S_{i,k}$ is formed by adding to $L_{i,k}$ a path $P_{i,k}(v)$ for each endpoint $v$ of an edge in $X$, where $P_{i,k}(v)\subset T$ connects $v$ to the first member of $L_{i,k}$ on the path from $v$ to the root of $T$.

\begin{lemma}
\label{leveled-cutsets}
For any $i$ and $k$,
$G\setminus  S_{i,k}$ is planar.
\end{lemma}

\begin{proof}
Extend the depth function on vertices to a continuous function
$\lambda:\mathcal M\mapsto\Real$ that maps each vertex to its depth,
maps each edge of $G$ one-to-one onto the interval between the depths of its endpoints,
and maps each cell of $\mathcal M$ in such a way that the map has no local minima or maxima in the interior of the cell.
Partition the surface underlying the map $\mathcal M$ into submanifolds with boundary, by removing all points $x$ for which $\lambda(x)=i\bmod k$.  Then each submanifold is formed from $\mathcal M$ by cutting certain curves at depth $\lambda(x)=i+ak$ (for some integer $a$) and certain other curves
at depth $\lambda(x)=i+(a+1)k$.  If we cap off these cut curves with disks, we form a manifold $\mathcal M_a$ without boundary.  Within each of the added disks at depth $\lambda(x)=i+ak$, add a vertex, connected by edges to all the vertices on the boundary of the disk; the result is a graph $G_a$ embedded in $\mathcal M_a$, such that each edge and vertex in $G\setminus  S_{i,k}$ belongs to exactly one of the graphs $G_a$.  If we let $T_a$ be formed from the union of $T\cap G_a$ with the newly added edges in $G_a$, then $T_a$ is a forest disjoint from the coforest $C\cap G_a$,
so we can extend these two subgraphs to a tree-cotree decomposition $(T'_a,C'_a,X'_a)$
such that $X'_a\subset X\cap G_a$.  Therefore, the paths $P_{i,k}(v)$ for $v$ an endpoint of an edge in $X\cap G_a$ form a cutset for $\mathcal M_a$, and $S_{i,k}$ is a cutset for all of $\mathcal M$.
\end{proof}

\begin{lemma}
\label{combined-cutset-size}
For any $k$, and any map embedded on a surface with Euler characteristic $\chi$,
$$\sum_{0\le i<k} |S_{i,k}|\le n+(\chi+2)k(k-1).$$
\end{lemma}

\begin{proof}
Each vertex of the map belongs to exactly one graph $L_{i,k}$.
In addition, each of the $2(\chi+2)$ endpoints $v$ of edges in $X$
contributes $\sum |P_{i,k}(v)|=\sum_{0\le i<k}i$ vertices to the total.
Some vertices may be in more than one path $P_{i,k}(v)$ but this only reduces the total.
\end{proof}

\begin{theorem}
Any $n$-vertex graph $G$ embedded on a surface with Euler characteristic $\chi$ has an $\epsilon$-separator
with $|S|\le\sqrt{(4\chi+O(1/\epsilon))n}$ vertices.
\end{theorem}

\begin{proof}
By choosing $i$ minimizing $|S_{i,k}|$,
we achieve $|S_{i,k}|\le n/k + \chi k + O(k)$.
The result follows by choosing $k=\sqrt{n/\chi}$, applying a planar separator theorem to the planar graph $G\setminus S_{i,k}$, and forming the union of the planar separator with $S_{i,k}$.
\end{proof}

If a map representing a characteristic-$\chi$ embedding of $G$ is given as input, the method described in the proof of the theorem can be implemented to run in time linear in the size of the map, and produces a cutset of size at most $\sqrt{(4\chi+O(1)) n}$.  The time bound for the overall algorithm, and the constant factor in the $\sqrt{O(n/\epsilon)}$ term in the separator size, depend on the choice of planar separator algorithm used to complete the separator construction.

We note that not all loops need always be incorporated in $S_{i,k}$ in order to make $G\setminus S_{i,k}$ planar. For instance, if $G$ is embedded on a torus, it can be cut into a planar graph in the form of a cylinder by removing only one of the two loops.  In addition, if we choose $S_{i,k}$ in such a way that some of the subgraphs $T_a$ and $C\cap G_a$ described in the proof of Lemma~\ref{leveled-cutsets} are disconnected, then additional edges are removed from $X$ in forming the tree-cotree decompositions of $\mathcal M_a$ and again we can reduce the number of loops. Can we use these observations to further reduce the constant factor in our separator theorem?

\section{Low-Diameter Tree Decomposition}

In previous work~\cite{Epp-Algo-00} we showed that in a minor-closed graph family, the treewidth of a graph with diameter $D$ can be bounded by a function of $D$, if and only if the family excludes some apex graph.  For instance, for planar graphs (which exclude the apex graph $K_5$) the treewidth is at most $3D-O(1)$~\cite{Bak-JACM-94}.
Such {\em bounded local treewidth} results can be used to find efficient solutions to many important algorithmic problems on these graphs~\cite{Bak-JACM-94,Epp-JGAA-99,Epp-Algo-00,math.CO/0001128}.

In particular, any graph embedded on a genus $g=O(1)$ surface has bounded local treewidth, since the family of such graphs is minor-closed and excludes an apex graph.  However, the treewidth bound from this general result grows rapidly with $g$, and in this special case we were able to prove a tighter bound:

\begin{theorem}[Eppstein~\cite{Epp-Algo-00}]
Let $G$ have genus $g$ and diameter $D$.
Then $G$ has treewidth $O(gD)$.
\end{theorem}

The proof relies on showing the existence of a cutset of size $O(gD)$, and forming a tree-decomposition by adding this cutset to each node in a tree-decomposition of the remaining planar graph.  However, one step in the proof is to find the minimum cardinality cutset, so the proof is nonconstructive, and Erickson and Sar-Peled~\cite{EriHar-SCG-02} later showed that in fact finding minimum cutsets is NP-complete.  By replacing this step with methods based on our tree-cotree decomposition, we can turn this nonconstructive proof into an efficient algorithm:

\begin{theorem}
Let $G$ have genus $g$ and diameter $D$, and suppose that we are given as input an embedding $\mathcal M$ of $G$ on a surface of genus $g$.
Then in time $O(gDn)$ we can find a tree-decomposition of $G$ with treewidth $O(gD)$.
\end{theorem}

\begin{proof}
We form a tree-decomposition $(T,C,X)$ by letting $T$ be a breadth-first spanning tree of $G$,
and letting $C$ be any spanning cotree disjoint from $T$.
We then let $S$ be the set of vertices incident to the edges in loops $\{{\rm loop}(T,e)\mid e\in X\}$.
Each loop has at most $2D$ edges, so $|S|=O(gD)$, and $G\setminus S$ is planar, so it has a tree-decomposition of treewidth $3D-O(1)$ that can be found in time~$O(Dn)$~\cite{Bak-JACM-94}.
We form a tree-decomposition of $G$ by adding $S$ to each node in the tree-decomposition of $G\setminus S$.
\end{proof}

The $O(gDn)$ time bound is optimal in the sense that it is the size of an explicit representation of the resulting tree-decomposition, but if we represent the sets of vertices at each tree-decomposition node implicitly, as a union of a constant number of paths in the two trees $T$ and $B$, the time can be reduced to linear in the size of the input map $\mathcal M$.

\section{Conclusions}

We have shown how the tree-cotree decomposition can be used in several applications for graphs embedded on surfaces: maintaining dynamic properties of the graph, computing separators, and constructing tree-decompositions of small treewidth.  There are many other important planar graph algorithms that can be generalized to graphs embedded on surfaces, and it would be interesting to see which others of these can be improved by the use of our tree-cotree decomposition methods.

\section*{Acknowledgements}

This research was supported in part by NSF grant CCR-9912338.

\raggedright
\bibliographystyle{abuser}
\bibliography{topgraph}

\begin{thebibliography}{10}

\bibitem{AleDji-SJDM-96}
L.~Aleksandrov and H.~Djidjev.
\newblock {Linear algorithms for partitioning embedded graphs of bounded
  genus}.
\newblock {\em SIAM J. Discrete Math.} 9:129--150, 1996,
  \url{http://www.dcs.warwick.ac.uk/~hristo/papers/genus-separator.ps.Z}.

\bibitem{Bak-JACM-94}
B.~S. Baker.
\newblock {Approximation algorithms for NP-complete problems on planar graphs}.
\newblock {\em J. Assoc. Comput. Mach.} 41:153--180, 1994.

\bibitem{BonLit-95}
C.~P. Bonnington and C.~H.~C. Little.
\newblock {\em {The Foundations of Topological Graph Theory}}.
\newblock Springer-Verlag, 1995.

\bibitem{Epp-JGAA-99}
D.~Eppstein.
\newblock {Subgraph isomorphism in planar graphs and related problems}.
\newblock {\em J. Graph Algorithms {\&} Applications} 3(3):1--27, 1999,
  arXiv:cs.DS/9911003.

\bibitem{Epp-Algo-00}
D.~Eppstein.
\newblock {Diameter and treewidth in minor-closed graph families}.
\newblock {\em Algorithmica} 27:275--291, 2000,
  \url{http://link.springer-ny.com/link/service/journals/00453/contents/00/100%
20/}, arXiv:math.CO/9907126.

\bibitem{EppGalIta-JCSS-96}
D.~Eppstein, Z.~Galil, G.~F. Italiano, and T.~H. Spencer.
\newblock {Separator based sparsification I: planarity testing and minimum
  spanning trees}.
\newblock {\em J. Computing {\&} Systems Sciences} 52(1):3--27, February 1996.

\bibitem{EppItaTam-Algs-92}
D.~Eppstein, G.~F. Italiano, R.~Tamassia, R.~E. Tarjan, J.~R. Westbrook, and
  M.~Yung.
\newblock {Maintenance of a minimum spanning forest in a dynamic planar graph}.
\newblock {\em J. Algorithms} 13(1):33--54, March 1992.

\bibitem{EriHar-SCG-02}
J.~Erickson and S.~Har-Peled.
\newblock {Optimally cutting a surface into a disk}.
\newblock {\em Proc. 18th ACM Symp. Computational Geometry}, pp.~244--253, June
  2002, \url{http://compgeom.cs.uiuc.edu/~jeffe/pubs/schema.html},
  arXiv:cs.CG/0207004.

\bibitem{math.CO/0001128}
M.~Grohe.
\newblock {Local tree-width, excluded minors, and approximation algorithms}.
\newblock arXiv.org e-Print archive, January 2000, arXiv:math.CO/0001128.

\bibitem{GroTuc-01}
J.~L. Gross and T.~W. Tucker.
\newblock {\em {Topological Graph Theory}}.
\newblock Dover Publications, 2001.

\bibitem{HenKin-JACM-99}
M.~R. Henzinger and V.~King.
\newblock {Randomized fully dynamic graph algorithms with polylogarithmic time
  per operation}.
\newblock {\em J. ACM} 46(4):502--536, July 1999.

\bibitem{HolLicTho-JACM-01}
J.~Holm, K.~de~Lichtenberg, and M.~Thorup.
\newblock {Poly-logarithmic deterministic fully-dynamic graph algorithms for
  connectivity, minimum spanning tree, 2-edge, and biconnectivity}.
\newblock {\em J. ACM} 48(4):723--760, July 2001.

\bibitem{Lin-JCTB-82}
S.~Lins.
\newblock {Graph-encoded maps}.
\newblock {\em J. Combinatorial Theory, Ser. B} 32:171--181, 1982.

\bibitem{LipTar-SJAM-79}
R.~Lipton and R.~E. Tarjan.
\newblock {A separator theorem for planar graphs}.
\newblock {\em SIAM J. Applied Math.} 36(2):177--189, April 1979.

\bibitem{SleTar-JCSS-83}
D.~D. Sleator and R.~E. Tarjan.
\newblock {A data structure for dynamic trees}.
\newblock {\em J. Computing {\&} Systems Sciences} 26:362--391, 1983.

\bibitem{SpiTen-SCG-96}
D.~A. Spielman and S.-H. Teng.
\newblock {Disk packings and planar separators}.
\newblock {\em Proc. 12th ACM Symp. Computational Geometry}, pp.~349--358,
  1996, \url{http://www-math.mit.edu/~spielman/Research/separator.html}.

\bibitem{Sti-93}
J.~Stillwell.
\newblock {\em {Classical Topology and Combinatorial Group Theory}}.
\newblock Graduate Texts in Mathematics~72. Springer-Verlag, 2nd edition, 1993.

\bibitem{Tho-STOC-00}
M.~Thorup.
\newblock {Near-optimal fully-dynamic graph connectivity}.
\newblock {\em Proc. 32nd ACM Symp. Theory of Computing}, pp.~343--350, May
  2000.

\bibitem{VegYap-SCG-90}
G.~Vegter and C.-K. Yap.
\newblock {Computational complexity of combinatorial surfaces}.
\newblock {\em Proc. 6th ACM Symp. Computational Geometry}, pp.~102--111, 1990,
  \url{http://www.acm.org/pubs/citations/proceedings/compgeom/98524/p102-vegte%
r/}.

\end{thebibliography}

\end{document}